\documentclass[11pt]{article}
\usepackage{putex}
\usepackage{graphicx}
\usepackage{longtable}
\usepackage{latexsym,amsmath,amsfonts,amssymb}
\usepackage{bbm}
\usepackage{mathtools}
\usepackage{authblk}
\usepackage{hyperref}

\renewenvironment{thebibliography}[1]{%
\begin{oldthebibliography}{#1}%
\setlength\itemsep{-2mm}%
}%
{%
\end{oldthebibliography}%
}

\newcommand{\eg}{\textit{e.g.}}

\newcommand{\ie}{\textit{i.e.}}

\numberwithin{equation}{section}

\newcommand{\nn}{\nonumber}

\newcommand{\be}{\begin{equation}} \newcommand{\ee}{\end{equation}}
\newcommand{\bea}{\begin{equation} \begin{aligned}} \newcommand{\eea}{\end{aligned} \end{equation}}

\newcommand{\cN}{\mathcal{N}}

\DeclareMathOperator{\Tr}{Tr}

\newtheorem{conj}{Conjecture}
\newcommand{\arrow}{\item[$\rightarrow$]}

\begin{document}


\title{Chiral Algebras for Trinion Theories}

\authors{\textbf{Madalena Lemos and Wolfger Peelaers}\\
\vspace{10pt}
C.N. Yang Institute for Theoretical Physics, \\
 Stony Brook University,\\
 Stony Brook, NY, 11794}

\abstract{\noindent It was recently understood that one can identify a chiral algebra in any four-dimensional $\mathcal N=2$ superconformal theory. 
In this note, we conjecture the full set of generators of the chiral algebras associated with the $T_n$ theories. The conjecture is motivated by making manifest the critical affine module structure in the graded partition function of the chiral algebras, which is computed by the Schur limit of the superconformal index for $T_n$ theories. We also explicitly construct the chiral algebra arising from the $T_4$ theory. Its null relations give rise to new $T_4$ Higgs branch chiral ring relations. }

\preprint{YITP-SB-14-41}

\maketitle


{
\setcounter{tocdepth}{2}
\setlength\parskip{-0.7mm}
\tableofcontents
}

\section{Introduction and conclusions}
In recent papers \cite{Beem:2013sza,Beem:2014kka} it was shown that even-dimensional extended superconformal field theories (SCFTs)\footnote{More precisely $\cN=(2,0)$ in $d=6$, $\cN \geq 2 $ in $d=4$, and ``small'' $\cN =(0,4)$ and $\cN=(4,4)$ in $d=2$.} contain a protected subsector that is isomorphic to a two-dimensional chiral algebra. This subsector is obtained by restricting operators to be coplanar and treating them at the level of cohomology with respect to a particular nilpotent supercharge, obtained as a combination of a supercharge and a superconformal charge of the theory. 
In showing the existence of the chiral algebra one relies only on the symmetries of the theory and there is no need to have a Lagrangian description --- a fact that was used to study the chiral algebras associated with the six-dimensional $(2,0)$-theory in \cite{Beem:2014kka} and with those obtained from four-dimensional theories of class $\mathcal S$ in \cite{Beem:2014rza}.
In this note we will focus on the chiral algebras associated with the so-called trinion or $T_n$ theories of class $\mathcal S.$

Chiral algebras of class $\mathcal S,$ \ie, the collection of chiral algebras associated with four-dimensional theories of class $\mathcal S$ \cite{Gaiotto:2009we,Gaiotto:2009hg}, were argued to take the form of a generalized topological quantum field theory (TQFT) in \cite{Beem:2014rza}. Within this TQFT, gluing, the operation associated to four-dimensional exactly marginal gauging, is achieved by solving a BRST cohomology problem, and partially closing a puncture is implemented via a quantum Drinfeld-Sokolov reduction. Furthermore, just as the isolated, strongly interacting $T_n$ theories, \ie{}, the theories whose UV-curve is a sphere with three punctures of maximal type, are the basic building blocks of class $\mathcal S$ theories, so are their associated chiral algebras the basic building blocks of said TQFT. Characterizing the $T_n$ chiral algebras is thus a prerequisite for an in principle complete understanding of chiral algebras of class $\mathcal S.$

However, while the existence of a chiral algebra inside a generic $\mathcal N=2$ SCFT can be argued in general terms, a complete characterization of its generators is currently lacking.\footnote{For Lagrangian theories, this problem can (in principle) be circumvented by explicitly constructing the full chiral algebra from the basic known chiral algebras associated with the free hypermultiplet and vector multiplet.} As for a partial characterization, it was argued in \cite{Beem:2013sza} that one is guaranteed to have at least generators in one-to-one correspondence with the Higgs branch chiral ring generators.\footnote{More generally, in the terminology of \cite{Beem:2013sza}, all generators of the so-called Hall-Littlewood chiral ring give rise to generators of the chiral algebra. For class $\mathcal{S}$ theories with acyclic generalized quivers, such as the $T_n$ theories, the Hall-Littlewood chiral ring equals the Higgs branch chiral ring.} In particular, the $T_n$ Higgs branch chiral ring contains as generators three moment map operators, one for each factor in the $T_n$ flavor symmetry algebra $\bigotimes _{i=1}^{3}\mathfrak{su}(n)_i$, and it was shown in \cite{Beem:2013sza} that their corresponding chiral algebra generators are three affine currents with affine levels $k_{2d,i}$ determined in terms of the four-dimensional flavor central charges $k_{4d,i}$ as $k_{2d,i}=-\frac{k_{4d,i}}{2}.$ These central charges are equal for the three factors, $k_{4d,i}=2n$, and thus the affine current algebras $\widehat{\mathfrak{su}(n)}$ have critical level $k_{2d}\equiv k_{2d,i} =-n$. 
The remaining generators of the $T_n$ Higgs branch chiral ring give rise to additional generators of the chiral algebra, which must be primaries of the affine Kac-Moody (AKM) algebras. 

It was also shown in \cite{Beem:2013sza} that the existence of a four-dimensional stress tensor implies that the chiral algebra must contain a meromorphic stress tensor. Therefore the global $\mathfrak{sl}(2)$ conformal algebra enhances to a Virasoro algebra, with the central charge fixed in terms of the four-dimensional $c$-anomaly coefficient by $c_{2d}=-12 c_{4d}$. However, the stress tensor is not necessarily a new generator of the chiral algebra, as it could be a composite operator (\ie{}, obtained from normal-ordered products of the generators and of their derivatives). Since the AKM current algebras are at the critical level, they do not admit a normalizable Sugawara stress tensor, and therefore the stress tensor can only be a composite if additional dimension two singlet composites can be constructed. This is only possible (and in fact happens) for $n=2$ and $3$.

In the first part of this note we perform a detailed study of the graded partition function of the $T_n$ chiral algebra, which can be computed thanks to its equality to the so-called Schur limit of the $\mathcal N=2$ superconformal index \cite{Gadde:2011ik,Gadde:2011uv}, and which shows that the collection of generators listed so far is not complete for $n>4$ (see section \ref{section_index}). Motivated by this analysis, we conjecture the complete set of generators to be as follows:
\begin{conj}[$T_n$ chiral algebra]
The $T_n$ chiral algebra $\chi(T_n)$ is generated by
\begin{itemize}
\item The set of operators, $\mathcal{H}$, arising from the Higgs branch chiral ring:
\begin{itemize}
\arrow Three $\widehat{\mathfrak{su}(n)}$ affine currents $J^1, J^2, J^3,$ at the critical level $k_{2d}=-n$, one for each factor in the flavor symmetry group of the theory,
\arrow Generators $W^{(k)}$, $k=1,\,,\ldots,\,n-1$ in the $(\wedge^k,\wedge^k,\wedge^k)$ representation of $\bigotimes _{j=1}^{3}\mathfrak{su}(n)_j$, where $\wedge^k$ denotes the $k-$index antisymmetric representation of $\mathfrak{su}(n)$. These generators have dimensions $\frac{k(n-k)}{2}$,
\end{itemize}
\item Operators $\mathcal O_i$, $i=1,\ldots n-1,$ of dimension $h_i=i+1$ and singlets under $\bigotimes _{j=1}^{3}\mathfrak{su}(n)_j$, with the dimension $2$ operator corresponding to the stress tensor $T$ of central charge $c_{2d}=-2n^3 + 3n^2 + n - 2$,
\end{itemize}
modulo possible relations which set some of the operators listed above equal to composites of the remaining generators.
\label{Tnconjecture}
\end{conj}

In other words, if one were to start with all generators of the above conjecture, one would find that some of them could be involved in null relations with composite operators, thereby being redundant. For example, in the chiral algebra associated with $T_2,$ \ie{}, the theory of eight free half-hypermultiplets, the affine currents and the stress tensor can be written as composites of the dimension $\frac{1}{2}$ generator $W^{(1)}$. For the case of $T_3$, which corresponds to the $E_6$ theory of \cite{Minahan:1996fg}, convincing evidence was provided in \cite{Beem:2014rza} that its chiral algebra $\chi(T_3)$ is fully generated by operators originating from the Higgs branch chiral ring. The stress tensor can be written as a composite and also, although not explicitly constructed in \cite{Beem:2014rza}, the dimension three singlet operator is accounted for as a composite. For $n>3$, as argued above, the stress tensor cannot be a composite of generators in $\mathcal{H}$, but the remaining dimension $3,\ldots, n$ singlet generators could still be. In the case of the $T_4$ chiral algebra the dimensions three and four singlet generators are redundant, as will be shown in section~\ref{section_T4}.

Our aim in the second part of this note is to verify Conjecture \ref{Tnconjecture} for $T_4$, in which case the chiral algebra is generated by the operators in $\mathcal{H}$ and the stress tensor, by explicitly constructing an associative algebra with these generators. Our approach to bootstrap this problem is to write down the most general operator product expansions (OPEs) between the generators, and to demand associativity of the operator product algebra by imposing the Jacobi-identities. Since chiral algebras are very rigid, one can hope that these constraints are sufficiently stringent to completely fix the operator algebra, as was famously shown to be the case for the first time for the $\mathcal{W}_3$ algebra in \cite{Zamolodchikov:1985wn} (see for example \cite{Bouwknegt:1992wg} for a review of other cases). We indeed find that the OPEs are completely and uniquely fixed. The analysis of the Jacobi-identities becomes technically involved in several instances, and as a result we can only claim that the conditions analyzed are necessary for an associative operator product algebra. However we believe that the remaining Jacobi-identities provide redundant constraints. As an interesting by-product of the explicit $T_4$ chiral algebra, we can compute four-dimensional Higgs branch chiral ring relations, which appear as null relations in the chiral algebra setting. Some of these relations are already known in the literature, (\eg{}, \cite{Benini:2009mz,Maruyoshi:2013hja}), and recovering them here provides a further check of the chiral algebra, while others are new.\\

As mentioned, four-dimensional Higgs branch chiral ring relations can be obtained from null relations in the chiral algebra. The explicit construction of $\chi(T_4)$ we present here thus provides a new, conceptually clear method to obtain all Higgs branch chiral ring relations for the $T_4$ theory. It seems plausible that once their structure is understood, they can be generalized to arbitrary $T_n.$ In this note we obtain all $\chi(T_4)$ null relations of dimension smaller than four, already uncovering new Higgs branch chiral ring relations, but the procedure can be taken further. For example, it would be possible to verify the recently proposed null relation of \cite{Hayashi:2014hfa}, as well as uncover further unknown ones. Furthermore, as will be elaborated upon in the next sections, our interpretation of the $\chi(T_n)$ chiral algebra partition function also predicts the existence of certain types of null relations, facilitating the task of explicitly computing them in the chiral algebra setting. 

Further checks of the $\chi(T_4)$ chiral algebra could be performed by partially closing punctures (via a quantum Drinfeld-Sokolov (qDS) reduction (see \cite{Beem:2014rza})) to obtain the free hypermultiplet, the $E_7$ theory of \cite{Minahan:1996cj}, or more generally the other fixtures of \cite{Chacaltana:2010ks}. For example, the chiral algebra associated with the $E_7$ theory is conjectured to be described by an affine $\widehat{\mathfrak e_7}$ current algebra at level $k_{2d}=-4$ and it is easy to convince oneself that the qDS procedure associated with the relevant $\mathfrak {su}(2)$ embedding will indeed result in dimension one currents corresponding to the decomposition of the ${\mathfrak e_7}$ adjoint representation. As shown in \cite{Beem:2014rza}, to complete the reduction argument, certain null relations need to exist in order to remove redundant generators in the reduced algebra. Such null relations are expected to descend from those of $\chi(T_4).$ 

The construction of $\chi(T_4)$ here made use of the constraints arising from associativity of the operator algebra. It would also be interesting to study if the theory space bootstrap, as introduced in \cite{Beem:2014rza}, which imposes instead associativity of the TQFT structure, might result in a complementary route to construct the chiral algebra. In particular with an eye towards a construction of $\chi(T_n),$ for $n>4$, an alternative (or a combined) approach might prove useful.\\

The organization of this note is as follows. In section~\ref{section_index} we analyze the  partition function of $\chi(T_n)$ employing its equality to the superconformal index of $T_n$ theories, and show how it motivates Conjecture \ref{Tnconjecture}, as well as some other expectations about the chiral algebra. In section~\ref{section_T4} we present the explicit construction of the $T_4$ chiral algebra and give explicit expressions for various null relations. We also show how our expectations deduced from the superconformal index are realized for $T_4$. The readers interested only in the explicit construction of $\chi(T_4)$ can safely skip section~\ref{section_index} as section~\ref{section_T4} is mostly independent from it. Appendix~\ref{App:character} contains some technical details on the relation between critical affine characters and the superconformal index, and in appendix~\ref{App:OPE} we collect all singular OPEs defining the chiral algebra $\chi(T_4).$

\section{$T_n$ indexology}
\label{section_index}

In this section we analyze the partition function of the $T_n$ chiral algebra, which gives insights into its generators and relations. By writing the partition function in a suggestive way we can justify Conjecture~\ref{Tnconjecture} and infer some properties of the structure of the chiral algebra, such as its null relations.

As shown in \cite{Beem:2013sza}, the graded partition function of the chiral algebra $\chi(T_n)$ equals the so-called Schur limit of the superconformal index of the $T_n$ theory \cite{Gadde:2011ik,Gadde:2011uv}. We work under the assumption that all generators are bosonic and thus the grading is immaterial. In appendix \ref{App:character} we show that the index can be written in a way suggestive of its interpretation as a two-dimensional partition function as
\begin{equation}
\mathcal{Z}_{\chi(T_n)}(q; \mathbf x_i) = \sum_{\mathfrak{R}_{\lambda}}  q^{\langle \lambda, \rho \rangle}  C_{\mathfrak{R}_{\lambda}}(q) \prod_{i=1}^{3}\text{ch}_{\mathfrak R_{\lambda}}(q,\mathbf x_i)\,.
\label{partitionfunction}
\end{equation}
Here $\mathbf{x}_i$ denote flavor fugacities conjugate to the Cartan generators of the $\mathfrak{su}(n)_i$ flavor symmetry associated with each of the three punctures, and the sum runs over all irreducible $\mathfrak{su}(n)$ representations $\mathfrak R_\lambda$ of highest weight $\lambda.$ The summand contains the product of three copies --- one for each puncture --- of $\text{ch}_{\mathfrak R_\lambda}(q,\mathbf x),$ the character of the critical irreducible highest weight representation of the affine current algebra $\widehat{\mathfrak{su}(n)}_{-n}$ with highest weight $\hat \lambda$, whose restriction to $\mathfrak{su}(n)$ is the highest weight $\lambda$ \cite{2007arXiv0706.1817A}.\footnote{Our notation here and in appendix \ref{App:character} follows that of \cite{DiFrancesco:1997nk}.} Furthermore, $\rho$ is the Weyl vector and $\langle \cdot , \cdot \rangle$ denotes the Killing inner product. Finally, the structure constants $C_{\mathfrak{R}_\lambda}(q)$ can be written as
\be
C_{\mathfrak{R}_\lambda}(q) 
= \mathrm{P.E.}\left[2 \sum_{j=1}^{n-1} \frac{q^{d_j}}{1-q} +2 \sum_{j=1}^{n-1} \left( n-j \right) q^j - 2\sum_{j=2}^{n}\sum_{1\leq i < j} q^{\ell_i - \ell_j + j-i} \right]\,,
\label{structrureconst}
\ee
where  $\ell_{i=1,\ldots, n}$ denote the lengths of rows of the Young tableau describing representation $\mathfrak R_\lambda$ with $\ell_n=0$, $d_j$ are the degrees of invariants, \ie{} $d_j = j+1$ for $\mathfrak{su}(n),$ and finally $\mathrm{P.E.}$ denotes the standard plethystic exponential
\be 
\mathrm{P.E.}\left[f(x)\right] = \exp \left( \sum_{m=1}^\infty \frac{f(x^m)}{m}\right)\,.
\ee
Let us provide some preliminary interpretative comments:
\begin{itemize}
\item We have obtained an expression for the partition function \eqref{partitionfunction} that is manifestly organized in terms of modules of the direct product of the three critical affine current algebras $\bigotimes_{i=1}^3 \widehat{(\mathfrak{su}(n)_i)}_{-n}$. Indeed, the factor $ q^{\langle \lambda, \rho \rangle} \prod_{i=1}^{3}\text{ch}_{\mathfrak R_{\lambda}}(q,\mathbf x_i)$ in \eqref{partitionfunction} captures threefold AKM primaries of dimension $\langle \lambda, \rho \rangle,$ transforming in representations $(\mathfrak R_\lambda,\mathfrak R_\lambda,\mathfrak R_\lambda)$, including for example all the $W^{(k)},$ and all of their AKM descendants.
\item The role of the structure constants is to encode additional operators beyond those captured by the threefold AKM modules. In particular, in the term $\mathfrak{R}_{\lambda=0}$ in the sum over representations, the structure constant $C_{\mathfrak{R}_{\lambda=0}}(q) =  \mathrm{P.E.}\left[2 \sum_{j=1}^{n-1} \frac{q^{d_j}}{1-q}\right]$  encodes two sets of additional operators of dimensions $d_j=j+1$, for $j=1,\ldots,n-1$, (and their $\mathfrak{sl}(2)$ descendants) acting on the vacuum module. These operators can either be new generators, or obtained as singlet composites of the generators captured by the AKM modules, which themselves are not present in the modules. Let us now describe these two sets:
\begin{enumerate}
\item The fact that the three AKM current algebras are at the critical level implies that all the Casimir operators  $\Tr(J^1)^k,\ \Tr(J^2)^k,\  \Tr(J^3)^k$ with $k=2,3,\ldots,n$ are null within their respective AKM algebra, and therefore that their action is not included in the affine modules. However, these operators do not remain null in the full chiral algebra, as it contains a stress tensor as well. In fact, null relations set all Casimirs equal $\Tr(J^1)^k =\ \Tr(J^2)^k =  \Tr(J^3)^k$.\footnote{The existence of these null relations follows directly from the existence of relations on the Higgs branch chiral ring setting the Casimir operators formed out of the moment map operators of the three flavor symmetries equal \cite{Maruyoshi:2013hja}. The corresponding chiral algebra null relations will be recovered in the next section.} These  $n-1$ Casimirs correspond to the first set of operators reinstated by the structure constants.
\item The second set of operators motivates our conjecture that there can be extra generators $\mathcal O_i$ with precisely dimensions $h_i=d_i=i+1$. 
\end{enumerate}
\end{itemize}
A more detailed discussion of these statements, and the interpretation of the two remaining factors in \eqref{structrureconst} is given in the remainder of this section. Readers not interested in this technical analysis can safely skip the remainder of this section.

\paragraph{ The AKM modules\\}
Ignoring for a moment the structure constants, each term in the sum over representations $\mathfrak R_\lambda$ of \eqref{partitionfunction} captures the states in the direct product of three critical affine modules with primary transforming in representation $(\mathfrak R_\lambda,\mathfrak R_\lambda,\mathfrak R_\lambda)$. The dimension of the threefold AKM primary is implemented by the factor $q^{\langle \lambda, \rho \rangle}$, yielding
\begin{equation}
h_{(\mathfrak R_\lambda, \mathfrak R_\lambda, \mathfrak R_\lambda)}=\langle \lambda, \rho \rangle  = \sum_{i=1}^{n-1}\frac{n-(2i-1)}{2}\ell_i\,.
\end{equation}
These pairings of dimension and representations include all the threefold AKM primary generators $W^{(k)},\  k=1,\ldots, n-1$ in Conjecture \ref{Tnconjecture}. (Note that the currents themselves are AKM descendants of the identity operator and appear in the vacuum module.) We expect that the remaining threefold AKM primaries in the sum over $\mathfrak R_\lambda$ all arise from combinations of normal-ordered products of generators in $\mathcal{H}$ (the set of generators originating from the Higgs branch chiral ring generators), and do not give rise to additional generators. It is clear that for each representation $(\mathfrak R_\lambda,\mathfrak R_\lambda,\mathfrak R_\lambda)$ one can write down a composite operator of the $W^{(k)}$, transforming in such representation, and with the appropriate dimension. Then, it seems plausible that such operator can always be made into a threefold AKM primary by --- if necessary --- adding composites of the remaining operators in $\mathcal{H}$. We have checked this statement in a few low-dimensional examples for $T_4$ (see equation \eqref{151515AKM} for an explicit example). 
All in all, the AKM modules capture the generators $W^{(k)}$, as well as other threefold AKM primaries obtained as their normal-ordered product, and all of their AKM descendants.

\paragraph{The structure constants\\}

The structure constants \eqref{structrureconst} encode additional operators on top of those captured by the AKM modules already described. Let us start by analyzing the factor
\be
\mathrm{P.E.}\left[2\sum_{j=1}^{n-1} \frac{q^{d_j}}{1-q}\right]\,.
\label{newgensandcas}
\ee
When inserted in \eqref{partitionfunction}, it encodes two sets of operators of dimensions $d_j$ and their $\mathfrak{sl}(2)$  descendants (taken into account by the denominator $\frac{1}{1-q}$), normal-ordered with all operators in any given AKM module $(\mathfrak R_\lambda,\mathfrak R_\lambda,\mathfrak R_\lambda)$. As described before, one set adds back the Casimir operators $\Tr(J^1)^k =\ \Tr(J^2)^k =  \Tr(J^3)^k$ of the AKM algebras,\footnote{Both the Casimirs and the Casimirs normal-ordered with threefold AKM primaries are new threefold AKM primaries, since they were null if one were to consider each AKM current algebra in isolation.} and the second set motivates the claim that there can be additional generators $\mathcal{O}_{i=1,\ldots,n-1}$ of dimensions $h_i=d_i=i+1$.\footnote{For readers familiar with the classification of four-dimensional superconformal multiplets of \cite{Dolan:2002zh}, these generators arise from four-dimensional operators in the $\hat{\mathcal{C}}$ multiplets.} However, one should bear in mind that in some cases one can construct (non-null) non-AKM-descendant singlet operators as composites of the $W^{(k)}$ of dimensions $h$ equal to one of these dimensions.
Since the only singlet operator in the sum over AKM modules, which is not an AKM descendant, corresponds to the identity operator, such operators must be accounted for by \eqref{newgensandcas}. This leaves two possibilities: it is either equal (or set equal by a null relation) to a composite of smaller dimensional $\mathcal{O}_{i}$ operators and/or of Casimirs, and consequently taken into account by the plethystic exponentiation in \eqref{newgensandcas}. Or it must take the place of the would-be generator $\mathcal{O}$ of dimension $h$.
In other words, if one were to include $\mathcal{O}$, one would find a null relation between this would-be generator and the composite of $W^{(k)}$.
As was mentioned before, the simplest example is the stress tensor $T\equiv \mathcal{O}_1$, which for $T_2$ and $T_3$ is a composite, but for $T_{n \geq 4}$ must be a new generator. In the next section we will show that for $T_4$ the generators of dimension three and four are absent, as the type of composites described above exist. However for $n \geq 5$ it is not possible to write such a composite of dimension three, and $\mathcal{O}_{2}$ must be a generator.

We now turn to the next factor in the structure constants \eqref{structrureconst}
\be
\mathrm{P.E.}\left[2\sum_{k=1}^{n-1} \left( n-k \right) q^k\right]\,.
\label{negativemodes}
\ee
Recalling that at the critical level the stress tensor is not obtained from the Sugawara construction, the critical modules do not contain derivatives\footnote{\ie{}, the action of the mode $L_{-1}$. As is common practice we use the mode expansion $\mathcal{O}(z)=\sum_n \frac{\mathcal{O}_n}{z^{n+h}}$ of an operator $\mathcal{O}$ of dimension $h$,
and $L_{n}$ denotes the modes of the stress tensor $T$.} of the threefold AKM primaries, although the full chiral algebra must. Similarly, the action of the modes $\left(\mathcal{O}_i\right)_{-1}$, $i=2,\ldots, n-1$,  $\left(\mathcal{O}_i\right)_{-2}$, $i=2,\ldots, n-1$, $\left(\mathcal{O}_i\right)_{-3}$, $i=3,\ldots, n-1$, $\ldots$ of the remaining singlet operators of the chiral algebra are not yet included. The number of modes we have to take into account at grade $k$ is precisely given by $n-k$, and these modes are added by one of the factors in \eqref{negativemodes}. The other factor adds back similar modes of the Casimir operators of the AKM current algebras, an explicit example of which will be given in the next section (see \eqref{JWCas}). It is clear that these modes cannot be added for all representations: for example, they cannot be added when considering the vacuum, since it is killed by all of them. Similarly, (and here we restrict to $n>2$) the only grade one modes acting on any of the $W^{(k)}$ that do not kill it must be the ones which correspond to either acting on it with a derivative, or normal-ordering it with a current, since these are the only ways one can write a dimension $\frac{k(n-k)}{2}+1$ composite in representation $(\wedge^k,\wedge^k,\wedge^k)$.\footnote{Note that the normal-ordered product $(JW^{(k)})$ in representation $(\wedge^k,\wedge^k,\wedge^k)$ is absent in the critical module.} These facts are taken into account by the factor 
\be 
\mathrm{P.E.}\left[- 2\sum_{j=2}^{n}\sum_{1\leq i < j} q^{\ell_i - \ell_j + j-i}\right]\,,
\label{nullsfornegativemodes}
\ee
which must subtract such relations, as well as other possible relations specific of each representation. Indeed, it is for example easy to verify that \eqref{negativemodes} and \eqref{nullsfornegativemodes} cancel each other for the vacuum module. For representations $(\wedge^k,\wedge^k,\wedge^k)$ only two $q$ terms survive in the plethystic exponential in the product of \eqref{negativemodes} and \eqref{nullsfornegativemodes}, which means that we are left with two grade one modes. One might have expected four grade one modes: one corresponding to acting with a derivative and three to normal-ordering with the three currents, but, as we will see in the next section, normal-ordering the three currents with $W^{(k)}$ (making an operator in representation $(\wedge^k,\wedge^k,\wedge^k)$) results in equal operators up to nulls (see equations \eqref{JW^(k) nulls} and \eqref{JWCas}).

As a final observation we note that the sum in \eqref{partitionfunction} only runs over flavor symmetry representations of the type $(\mathfrak R_\lambda,\mathfrak R_\lambda,\mathfrak R_\lambda)$, and the structure constants \eqref{structrureconst} cannot alter flavor symmetry information. Therefore the partition function predicts that any operator transforming in a representation $(\mathfrak R_{\lambda_1},\mathfrak R_{\lambda_2},\mathfrak R_{\lambda_3})$ with not all equal $\lambda_i$ cannot be a threefold AKM primary. More precisely, if we encounter an operator in unequal representations $(\mathfrak R_{\lambda_1},\mathfrak R_{\lambda_2},\mathfrak R_{\lambda_3})$ it must either be an AKM descendant, or obtained from one via the operators taken into account by the structure constants (namely by the action of any operators contributing to \eqref{newgensandcas} and \eqref{negativemodes}).
We will get back to this point in the next section (around example \eqref{null1066}).
\section{The $T_4$ chiral algebra}
\label{section_T4}

For the chiral algebra associated with the $T_4$ theory, Conjecture \ref{Tnconjecture} states that the collection of generators $\mathcal G$ contains three $\widehat{\mathfrak{su}(4)}$ affine currents at the critical level $k_{2d}=-4$, which we denote by $\left(J^1\right)^{b_1}_{a_1}$, $\left(J^2\right)^{b_2}_{a_2}$, $\left(J^3\right)^{b_3}_{a_3}$, two dimension $\frac{3}{2}$ generators, $W^{(1)}$ and $W^{(3)},$ in the tri-fundamental and tri-antifundamental representations of the flavor symmetry group respectively, which we rename $W_{a_1a_2a_3}$ and $\widetilde{W}^{b_1b_2b_3}$, and one dimension two generator, $W^{(2)},$ in the $\mathbf{6}\times\mathbf{6}\times\mathbf{6}$ representation which we denote explicitly as $V_{[a_1b_1][a_2b_2][a_3b_3]}$. Here $a_i, b_i, c_i,\ldots=1,2,3,4$ are (anti)fundamental indices corresponding to the flavor symmetry factor $\mathfrak{su}(4)_{i=1,2,3}$. Moreover, we must add the stress tensor $T$ as an independent generator, with central charge $c_{2d} = -78$, but we claim that the dimension three and four singlets operators can be obtained as composites. As will be shown later the dimension three operator is argued to be a Virasoro primary involving $W \widetilde{W}\vert_{\mathrm{sing}}$, where $\vert_{\mathrm{sing}}$ means we take the singlet combination, while the dimension four generator is a Virasoro primary combination involving $V V\vert_{\mathrm{sing}}$. We summarize the conjectured generators in Table \ref{table_T4generators}.
\begin{table}[h!]
\centering
\renewcommand{\arraystretch}{1.3}
\begin{tabular}{c|c|c}
generator $\mathcal G$ &  $\ h_{\mathcal G}\ $&  $\mathcal R_{\mathcal G}$\\
\hline\hline
$(J^1)_{b_1}^{a_1}$ & $1$ & ($\mathbf {15}$,$\mathbf {1}$,$\mathbf {1}$)\\
$(J^2)_{b_2}^{a_2}$ & $1$ & ($\mathbf {1}$,$\mathbf {15}$,$\mathbf {1}$) \\
$(J^3)_{b_3}^{a_3}$ & $1$ & ($\mathbf {1}$, $\mathbf {1}$,$\mathbf {15}$) \\
$T$& $2$ & ($\mathbf {1}$,$\mathbf {1}$,$\mathbf {1}$) \\
$W_{a_1a_2a_3}$& $\frac{3}{2}$ & ($\mathbf {4}$,$\mathbf {4}$,$\mathbf {4}$)\\
$\widetilde W^{a_1a_2a_3}$& $\frac{3}{2}$ & ($\mathbf {\bar{4}}$,$\mathbf {\bar{4}}$,$\mathbf {\bar{4}}$) \\
$V_{[a_1b_1][a_2b_2][a_3b_3]}$& 2 &($\mathbf {6}$,$\mathbf {6}$,$\mathbf {6}$)\\
\hline
\end{tabular}
\caption{$T_4$ generators $\mathcal G,$ their dimension $h_{\mathcal G}$ and their $\mathfrak{su}(4)^3$ representation $\mathfrak R_{\mathcal G}$.\label{table_T4generators}}
\end{table}

As mentioned before, our strategy for finding the $T_4$ chiral algebra is a concrete implementation of the conformal bootstrap program. 
We start by writing down the most general OPEs for this set of generators consistent with the symmetries of the theory, and in particular we impose that the three different flavor symmetry groups appear on equal footing. This of course implies that the three flavor currents have the same affine level, simply denoted by $k_{2d}$.
The OPEs of all the generators with the stress tensor are naturally fixed to be those of Virasoro primaries with the respective dimensions. Moreover, all generators listed in Table \ref{table_T4generators}, with the exception of the stress tensor\footnote{It is clear that the stress tensor cannot be an AKM primary, as the OPE between a dimension one operator (the current) and the stress tensor must have necessarily a $\frac{1}{z^2}$ pole. It is also not an AKM descendant, since at the critical level it cannot be given by the Sugawara construction.}, are affine Kac-Moody primaries of the three current algebras, transforming in the indicated representation. Therefore their OPEs with the currents are also completely fixed. In the self-OPEs of the AKM currents and the stress tensor, we could fix the affine level and the central charge to the values corresponding to the $\chi(T_4)$ chiral algebra, $k_{2d}=-4$ and $c_{2d}=-78$. Instead we leave them as free parameters and try to fix them the same way as any other OPE coefficient. For the remaining OPEs we write all possible operators allowed by the symmetries of the theory with arbitrary coefficients. Our expectation is that this chiral algebra is unique, and that by imposing associativity one can fix all the OPE coefficients, including $k_{2d}$ and $c_{2d}$. This indeed turns out to be true.  Some of the resulting OPEs are quite long so we collect them all in appendix \ref{App:OPE}.\footnote{In there and in what follows we adopt the standard conventions for the normal-ordering of operators such that $\mathcal{O}_1 \mathcal{O}_2\ldots \mathcal{O}_{\ell-1} \mathcal{O}_\ell = \left(\mathcal{O}_1 \left(\mathcal{O}_2 \ldots \left(\mathcal{O}_{\ell-1}\mathcal{O}_{\ell}\right)...\right)\right)$.}\\

The next step is to fix all the arbitrary coefficients by imposing Jacobi-identities, implementing in this way the requirement that the operator algebra is associative.
Concretely, we impose on any combination of three generators $A,B,C$ the Jacobi-identities (see, \eg{}, \cite{Thielemans:1994er})
\be 
\left[A(z) \left[ B(w) C(u) \right]\right]- \left[B(w)\left[ A(z) C(u)\right]\right]-\left[\left[ A(z) B(w))\right]C(u) \right]=0\,,
\ee
for $|w-u| < |z-u|$, where  $\left[A(z) B(w)\right]$ denotes taking the singular part of the OPE of $A(z)$ and $B(w)$, and where we already took into account that our generators are bosonic and no extra signs are needed.
It is important to note that the Jacobi-identities do not need to be exactly zero, but they can be proportional to null operators. Since null operators decouple, associativity of the algebra is not spoiled. For analyzing the Jacobi-identities we make use of the Mathematica package described in \cite{Thielemans:1991uw}. Even so, the analysis is quite cumbersome due to the large number of fields appearing in the OPEs and the necessity of removing null relations, especially so for the Jacobi-identities involving the generator $V$. These null relations are not known \textit{a priori}, therefore part of the task consists of obtaining all null operators at each dimension and in a given representation of the flavor symmetry. Due to these technical limitations we have only found necessary conditions for the Jacobi-identities to be satisfied, not sufficient ones. Nevertheless these conditions turn out to fix completely all the OPE coefficients, including the level and the central charge of the theory, meaning the chiral algebra with this particular set of generators is unique. After all coefficients are fixed, the remaining Jacobi-identities analyzed serve as a test on the consistency of our chiral algebra. We have checked a large enough set of Jacobi-identities to be convinced that the remaining ones will not give any additional constraints. If that is the case we have found an associative operator algebra with the same set of generators and the same central charges as conjectured for the $T_4$ chiral algebra. A further check that the chiral algebra we constructed corresponds indeed to the $T_4$ chiral algebra can be performed by comparing the partition function of the former to the one of the latter (which is nothing else than the superconformal index of $T_4$). Whereas in section \ref{section_index} we have exploited the index to motivate our claim about the full set of generators of the $T_4$ chiral algebra, in what follows we perform a partial check of the equality of the actual partition function of the constructed chiral algebra with the index by comparing the null states of the chiral algebra to the ones predicted by the superconformal index up to dimension $\frac{7}{2}$. Even if there were generators that we have missed in this analysis, the facts that the generators in our chiral algebra must be present, and that the chiral algebra we constructed is closed (assuming we have solved all constraints from the Jacobi-identities), imply that we have found a closed subalgebra of the full $T_4$ chiral algebra.\\

\begin{table}[h!]
\centering
\renewcommand{\arraystretch}{1.3}
\begin{tabular}{c|c|c||c|c|c}
 $\ h\ $ & mult. & $\mathcal R$ &  $\ h\ $ & mult. & $\mathcal R$\\
\hline\hline
$2$ & $2$ & ($\mathbf{1}$,$\mathbf{1}$,$\mathbf{1}$)                     &   $\frac{7}{2}$     & 10 &    ($\mathbf{4}$,$\mathbf{4}$,$\mathbf{4}$), ($\mathbf{\bar{4}}$,$\mathbf{\bar{4}}$,$\mathbf{\bar{4}}$)        \\
$\frac{5}{2}$ & $2$ & ($\mathbf{4}$,$\mathbf{4}$,$\mathbf{4}$), ($\mathbf{\bar{4}}$,$\mathbf{\bar{4}}$,$\mathbf{\bar{4}}$)                                                        &        & 2  &       ($\mathbf{36}$,$\mathbf{4}$,$\mathbf{4}$), ($\mathbf{\bar{36}}$,$\mathbf{\bar{4}}$,$\mathbf{\bar{4}}$)   , and perms.     \\
 $3$  & $4$ & ($\mathbf{1}$,$\mathbf{1}$,$\mathbf{1}$)                     &         & 1  &   ($\mathbf{20}$,$\mathbf{20}$,$\mathbf{4}$), ($\mathbf{\bar{20}}$,$\mathbf{\bar{20}}$,$\mathbf{\bar 4}$)   , and perms.    \\
  & $2$ & ($\mathbf{6}$,$\mathbf{6}$,$\mathbf{6}$)                         &         & 3  &     ($\mathbf{20}$,$\mathbf{4}$,$\mathbf{4}$), ($\mathbf{\bar{20}}$,$\mathbf{\bar{4}}$,$\mathbf{\bar 4}$)   , and perms.       \\ 
& $1$ & ($\mathbf{6}$,$\mathbf{6}$,$\mathbf{10}$), and perms.                                                   &         & &          \\
 & $3$ & ($\mathbf{15}$,$\mathbf{1}$,$\mathbf{1}$), and perms.                                &         & &           \\
 & $1$ & ($\mathbf{15}$,$\mathbf{15}$,$\mathbf{1}$),  and perms.                          &         & &        \\
 \hline
\end{tabular}
\caption{Quantum numbers and multiplicities of $T_4$ null operators up to dimension $\frac{7}{2}.$\label{table_nullopsT4}}
\end{table}

For practical purposes, it is useful to rewrite the partition function of the $T_n$ chiral algebra \eqref{partitionfunction} alternatively as
\begin{equation}
\label{indexasgensandnulls}
\mathcal{Z}_{\chi(T_n)}(q; \mathbf x_i)  = \mathrm{P.E.}\left[  \frac{1}{1-q}\sum_{\text{generators } \mathcal G} q^{h_{\mathcal G}} \chi^{\mathfrak{su}(n)^3}_{\mathcal{R}_\mathcal{G}}(\mathbf{x}_i)  \right]-\sum_{\text{nulls } \mathcal{N}} q^{h_{\mathcal N}}\chi^{\mathfrak{su}(n)^3}_{\mathcal{R}_\mathcal{N}}(\mathbf{x}_i) \,,
\end{equation}
in terms of a piece that describes the generators $\mathcal G$ of dimensions $h_\mathcal{G}$ transforming in representations $\mathcal{R}_{\mathcal G}$ of the $\mathfrak{su}(n)^3$ flavor symmetry and their $\mathfrak{sl}(2)$ descendants, and a term that subtracts off explicitly the null operators $\mathcal N$, of dimension $h_{\mathcal N}$ and in representation $\mathcal{R}_\mathcal{N}$. By comparing the expansion in powers of $q$ of \eqref{partitionfunction} with that of \eqref{indexasgensandnulls} (and under the assumption that the full list of generators is as in Table \ref{table_T4generators}) we can predict how many nulls to expect in each representation and at each dimension. In Table \ref{table_nullopsT4} we summarize the resulting quantum numbers of the low-lying null operators $\mathcal N.$ We have explicitly constructed the null operators corresponding to the entries in Table \ref{table_nullopsT4}; the full list is given in Tables \ref{Nullrelations1} and \ref{Nullrelations2}, where we have defined $S^i$ to be the quadratic Casimir $S^i = (J^i)_{a_i}^{b_i}(J^i)_{b_i}^{a_i}$. 

\begin{table}[h]
\renewcommand{\arraystretch}{1.75}
\centering
\begin{tabular}{|c|c|l|}
\hline \hline 
\ $h_{\mathcal N}$\  &\  $\mathcal R_{\mathcal N}$ \ & \multicolumn{1}{|c|}{Null relations}\\
\hline
2 & $(\mathbf 1,\mathbf 1,\mathbf 1)$ & $(J^1)_{a_1}^{b_1}(J^1)_{b_1}^{a_1} = (J^2)_{a_2}^{b_2}(J^2)_{b_2}^{a_2}  =(J^3)_{a_3}^{b_3}(J^3)_{b_3}^{a_3} $\\
\hline

5/2 & $(\mathbf 4,\mathbf 4,\mathbf 4)$ & $(J^1)^{b_1}_{a_1}W_{b_1 a_2a_3}=(J^2)^{b_2}_{a_2}W_{a_1 b_2 a_3}=(J^3)^{b_3}_{a_3}W_{a_1 a_2 b_3}$ \\
\hline

3 & $(\mathbf 1,\mathbf 1,\mathbf 1)$ & $(J^1)_{a_1}^{b_1}(J^1)_{b_1}^{c_1}(J^1)_{c_1}^{a_1} = (J^2)_{a_2}^{b_2}(J^2)_{b_2}^{c_2}(J^2)_{c_2}^{a_2} = (J^3)_{a_3}^{b_3}(J^3)_{b_3}^{c_3}(J^3)_{c_3}^{a_3} $\\

&&$\partial\left((J^1)_{a_1}^{b_1}(J^1)_{b_1}^{a_1}\right)= \partial\left((J^2)_{a_2}^{b_2}(J^2)_{b_2}^{a_2}\right)=\partial\left((J^3)_{a_3}^{b_3}(J^3)_{b_3}^{a_3}\right)$ \\
\hline

3 & $(\mathbf 6,\mathbf 6,\mathbf 6)$ & $(J^1)_{[a_1}^{c_1} V_{[b_1]c_1][a_2b_2][a_3b_3]}=(J^2)_{[a_2}^{c_2} V_{[a_1b_1][b_2]c_2][a_3b_3]}=(J^3)_{[a_3}^{c_3} V_{[a_1b_1][a_2b_2][b_3]c_3]} $\\
\hline

3 & $(\mathbf {10},\mathbf 6,\mathbf 6)$ & $W_{(a_1[a_2[a_3}W_{b_1)b_2]b_3]} =-\frac{1}{4} J_{(a_1}^{c_1}V_{[|c_1|b_1)][a_2b_2][a_3b_3]} $\\
\hline

3 & $(\mathbf {15},\mathbf 1,\mathbf 1)$ &   $(J^1)_{a_1}^{b_1}(J^1)_{c_1}^{d_1}(J^1)_{d_1}^{c_1} = (J^1)_{a_1}^{b_1}(J^2)_{a_2}^{b_2}(J^2)_{b_2}^{a_2}  =(J^1)_{a_1}^{b_1}(J^3)_{a_3}^{b_3}(J^3)_{b_3}^{a_3} $\\

&& $(W_{a_1 a_2 a_3}\tilde W^{b_1 a_2 a_3}-\text{trace}) = \frac{1}{16} (J^1)_{a_1}^{b_1}(J^3)_{a_3}^{b_3}(J^3)_{b_3}^{a_3}  + \frac{1}{16}  (J^1)_{a_1}^{b_1}T$\\
&& \qquad $ + \left[ ((J^1)_{a_1}^{c_1} \partial (J^1)_{c_1}^{b_1}-\text{trace}) + ((J^1)_{c_1}^{b_1} \partial (J^1)_{a_1}^{c_1}-\text{trace})\right]  $\\
&& \qquad $- \frac{1}{4} ((J^1)_{a_1}^{c_1}(J^1)_{c_1}^{d_1}(J^1)_{d_1}^{b_1} - \text{trace}) $\\
\hline

3 & $(\mathbf {15},\mathbf {15},\mathbf 1)$ & $(W_{a_1 a_2 a_3}\tilde W^{b_1 b_2 a_3}-\text{traces})  = \frac{1}{4} \left[(J^1)_{a_1}^{b_1}\partial(J^2)_{a_2}^{b_2} + (J^2)_{a_2}^{b_2}\partial(J^1)_{a_1}^{b_1}\right] $ \\
&& \qquad $-\frac{1}{16}  \left[((J^1)_{a_1}^{b_1}(J^2)_{a_2}^{c_2}(J^2)_{c_2}^{b_2} -\text{trace}) + ((J^2)_{a_2}^{b_2}(J^1)_{a_1}^{c_1}(J^1)_{c_1}^{b_1} -\text{trace})\right]$ \\
\hline
\end{tabular}
\caption{Explicit null relations up to dimension three, which can be uplifted to four-dimensional Higgs branch chiral ring relations. Representations which are not real give rise to a similar null in the complex conjugate representation,  and representations which are not equal in the three flavor groups give rise to similar null relations with permutations of the flavor group indices.  Note that these, togheter with Table~\ref{Nullrelations2}, are in one-to-one correspondence to the null relations subtracted from the index given in Table \ref{table_nullopsT4}. \label{Nullrelations1}}
\end{table}

Null relations in the two-dimensional chiral algebra can be uplifted to four-dimensional Higgs branch chiral ring relations, a partial list of which is given in \cite{Maruyoshi:2013hja}, by setting to zero all derivatives and generators not coming from the Higgs branch chiral ring (in particular the stress tensor and the other singlet generators, if present as independent generators). The nulls of Tables \ref{Nullrelations1} and \ref{Nullrelations2} allow one to recover the low-dimensional Higgs branch chiral ring relations in \cite{Maruyoshi:2013hja}, and to find additional ones. Let us give a few illustrative examples.

\begin{table}[h!]
\renewcommand{\arraystretch}{1.75}
\centering
\begin{tabular}{|c|c|l|}
\hline \hline 
\ $h_{\mathcal N}$\  &\  $\mathcal R_{\mathcal N}$ \ & \multicolumn{1}{|c|}{Null relations}\\
\hline

7/2 & $(\mathbf {20},\mathbf {4},\mathbf 4)$ & $ 8 \widetilde W^{f_1b_2b_3} V_{[d_1(e_1][a_2b_2][a_3b_3]}\ \epsilon_{|f_1|a_1)b_1c_1}$ \\
&&\qquad $=9(J^1)^{f_1}_{[a_1}(J^1)^{g_1}_{b_1}W_{(c_1]a_2a_3 }\ \epsilon_{d_1)e_1 f_1g_1 }  -2 (J^1)^{f_1}_{[d_1} (J^3)_{a_3}^{b_3}W_{(e_1]a_2b_3}\ \epsilon_{|f_1|a_1)b_1c_1} $ \\
&&\qquad $+ 3 \partial (J^1)_{[d_1}^{f_1} W_{(e_1]a_2a_3}\ \epsilon_{|f_1|a_1)b_1c_1} +6 \partial\left( (J^1)_{[d_1}^{f_1}  W_{(e_1]a_2a_3}\right)\ \epsilon_{|f_1|a_1)b_1c_1}$ \\

 & & $(J^1)^{f_1}_{[d_1} (J^2)_{a_2}^{b_2}W_{(e_1]b_2a_3}\ \epsilon_{|f_1|a_1)b_1c_1} = (J^1)^{f_1}_{[d_1} (J^3)_{a_3}^{b_3}W_{(e_1]a_2b_3}\ \epsilon_{|f_1|a_1)b_1c_1} $\\ 
 & & \qquad $= (J^1)^{f_1}_{[d_1} (J^1)_{(e_1]}^{g_1}W_{|g_1|a_2a_3}\ \epsilon_{|f_1|a_1)b_1c_1} $\\
\hline
7/2 & $(\mathbf {20},\mathbf {20},\mathbf {4})$ & $   \widetilde W^{f_1f_2b_3} V_{[d_1(e_1][d_2(e_2][a_3b_3]}\ \epsilon_{|f_1|a_1)b_1c_1}\ \epsilon_{|f_2|a_2)b_2c_2}$ \\
&&\qquad  $=-\frac{1}{2}(J^1)_{[d_1}^{f_1}(J^2)_{[d_2}^{f_2}W_{(e_1](e_2]a_3}\ \epsilon_{|f_1|a_1)b_1c_1}\ \epsilon_{|f_2|a_2)b_2c_2} $\\
\hline

7/2 & $(\mathbf {36},\mathbf {4},\mathbf {4})$ & $ (J^1)_{(a_1}^{b_1} (J^2)_{a_2}^{b_2} W_{c_1 b_2 a_3} \ \epsilon_{d_1) b_1 e_1 f_1}= 
		(J^1)_{(a_1}^{b_1} (J^3)_{a_3}^{b_3} W_{c_1 a_2 b_3} \ \epsilon_{d_1) b_1 e_1 f_1} = $ \\
	& & \qquad $(J^1)_{(a_1}^{b_1} (J^1)_{c_1}^{h_1} W_{|h_1| a_2 a_3} \ \epsilon_{d_1) b_1 e_1 f_1}$\\

\hline

7/2 & $(\mathbf {4},\mathbf {4},\mathbf {4})$ & $S^{1}W_{a_1a_2a_3}= S^2W_{a_1a_2a_3}=S^3W_{a_1a_2a_3}$\\

&& $8 \widetilde W^{b_1b_2b_3} V_{[b_1a_1][b_2a_2][b_3a_3]} = 2(J^1)_{a_1}^{b_1}(J^3)_{a_3}^{b_3}W_{b_1a_2b_3} +9  T W_{a_1a_2a_3}  $\\
&& \qquad $+15 \partial \left( (J^{3})_{a_3}^{b_3}  W_{a_1a_2b_3} \right) - \frac{9}{2}\partial^2 W_{a_1a_2a_3} - \frac{3}{2} S^{1}W_{a_1a_2a_3}$\\
&& \qquad $-9 \left((J^{1})_{a_1}^{b_1} \partial W_{b_1a_2a_3}+ (J^{2})_{a_2}^{b_2} \partial W_{a_1b_2a_3} +  (J^{3})_{a_3}^{b_3} \partial W_{a_1a_2b_3} \right) $\\

&& $(J^1)_{a_1}^{b_1}(J^2)_{a_2}^{b_2}W_{b_1b_2a_3} = (J^1)_{a_1}^{b_1}(J^3)_{a_3}^{b_3}W_{b_1a_2b_3} = (J^2)_{a_2}^{b_2}(J^3)_{a_3}^{b_3}W_{a_1b_2b_3}$\\
&& \qquad  $= (J^1)_{a_1}^{b_1}(J^1)_{b_1}^{c_1}W_{c_1a_2a_3}=(J^2)_{a_2}^{b_2}(J^2)_{b_2}^{c_2}W_{a_1c_2a_3}=(J^3)_{a_3}^{b_3}(J^3)_{b_3}^{c_3}W_{a_1 a_2 c_3}$\\

&& $\partial \left[(J^1)_{a_1}^{b_1}W_{b_1a_2a_3}\right]=\partial \left[ (J^2)_{a_2}^{b_2} W_{a_1b_2a_3} \right] = \partial \left[ (J^3)_{a_3}^{b_3} W_{a_1a_2b_3} \right] $\\
\hline

\end{tabular}
\caption{Explicit null relations at dimension $7/2$, which can be uplifted to four-dimensional Higgs branch chiral ring relations.  Representations which are not real give rise to a similar null in the complex conjugate representation,  and representations which are not equal in the three flavor groups give rise to similar null relations with permutations of the flavor group indices.  Note that these, togheter with Table~\ref{Nullrelations1}, are in one-to-one correspondence to the null relations subtracted from the index given in Table \ref{table_nullopsT4}. \label{Nullrelations2}}
\end{table}

A simple calculation shows that the null relations
\begin{equation}
\Tr(J^1)^2 = \Tr (J^2)^2 = \Tr (J^3)^2\,,
\end{equation}
hold true. Each of these operators separately is null within its respective critical current algebra, but thanks to the presence of the stress tensor $T$ in the full chiral algebra, one finds that only their differences are null. Similarly, we have explicitly recovered the analogous relation for the third order Casimir operators. These null relations are just two instances of the general null relations setting equal the Casimir operators of the three current algebras, which are similarly valid for general $T_n.$ The corresponding Higgs branch chiral ring relations on the moment map operators are well-known (see for example \cite{Maruyoshi:2013hja}).

Another nice set of null relations is obtained by acting with a current on the generators $W^{(k)}$:
\begin{align}
&(J^1)^{b_1}_{a_1}W_{b_1 a_2a_3}=(J^2)^{b_2}_{a_2}W_{a_1 b_2 a_3}=(J^3)^{b_3}_{a_3}W_{a_1 a_2 b_3}\,,\notag\\
&(J^1)_{b_1}^{a_1}\widetilde W^{b_1 a_2a_3}=(J^2)_{b_2}^{a_2}\widetilde W^{a_1 b_2 a_3}=(J^3)_{b_3}^{a_3}\widetilde W^{a_1 a_2 b_3}\,,\notag\\
&(J^1)_{[a_1}^{c_1} V_{[b_1]c_1][a_2b_2][a_3b_3]}=(J^2)_{[a_2}^{c_2} V_{[a_1b_1][b_2]c_2][a_3b_3]}=(J^3)_{[a_3}^{c_3} V_{[a_1b_1][a_2b_2][b_3]c_3]}\,.
\label{JW^(k) nulls}
\end{align}
Null relations of this type are expected to be valid in general $T_n$ as well, and extend the ones listed in \cite{Maruyoshi:2013hja} for $W^{(1)},W^{(n-1)}$.  Some of the null relations presented in Tables \ref{Nullrelations1} and \ref{Nullrelations2} are direct consequences of these nulls, obtained by either acting with derivatives or normal-ordering them with other operators, but others are new. For example, the last two nulls given in Table \ref{Nullrelations1} are not obtained from previous nulls, and they give rise to known Higgs branch chiral ring relations (they precisely turn into the relations given in equation (2.7) of \cite{Maruyoshi:2013hja} after setting all derivatives and the stress tensor to zero, and taking into account the different normalizations of the two- and four-dimensional operators). All null relations involving the generator $V$ in Tables \ref{Nullrelations1} and \ref{Nullrelations2} give rise to new Higgs branch chiral ring relations.

As mentioned, when computing Jacobi-identities one might find that some of them are not zero on the nose, but end up being proportional to null states. In practice this happens quite often, and we find that consistency of the Jacobi-identities relies precisely on the existence of some of these nulls. For example, when examining the Jacobi-identities involving $W$, $\widetilde{W}$ and $V$, one encounters the following null relation:
\be
W_{(a_1[a_2[a_3}W_{b_1)b_2]b_3]} = - \frac{1}{4} J_{(a_1}^{c_1}V_{[|c_1|b_1)][a_2b_2][a_3b_3]}\,,
\label{null1066}
\ee
which only exists at $k_{2d}=-4$.

We can now check a prediction made in section \ref{section_index}, namely that any operator transforming in a representation  $(\mathfrak{R}_{\lambda_1}, \mathfrak{R}_{\lambda_2}, \mathfrak{R}_{\lambda_3})$ for not all equal $\mathfrak{R}_{\lambda_i}$ must be an AKM descendant (or be obtained from an AKM descendant by acting on it with the operators which contribute to the structure constants). The operator $W_{(a_1[a_2[a_3}W_{b_1)b_2]b_3]}$ would seem to contradict this statement, since it transforms in the representation $(\mathbf{10},\mathbf{6},\mathbf{6})$, and it clearly cannot be obtained from an AKM descendant. Fortunately, there is no contradiction with the superconformal index as this operator is set equal to an AKM descendant by the null relation \eqref{null1066}.
More generally, we have verified in several cases that threefold AKM primaries either appear in representations of the type $(\mathfrak{R}_{\lambda}, \mathfrak{R}_{\lambda}, \mathfrak{R}_{\lambda})$, or are null. Moreover we have checked that all operators in representations which are not of the type $(\mathfrak{R}_{\lambda}, \mathfrak{R}_{\lambda}, \mathfrak{R}_{\lambda})$ are either AKM descendants or obtained from them by acting with the operators which contribute to the structure constants, such as a derivative, or normal-ordering it with the stress tensor.
A direct consequence of this interpretation of the partition function is that we can predict the existence of certain types of relations: whenever we can write an operator in a representation not of the type $(\mathfrak{R}_{\lambda}, \mathfrak{R}_{\lambda}, \mathfrak{R}_{\lambda})$ which is neither a descendant nor obtained from one in the manner described above, there has to be a null relation involving it. Since null operators are threefold AKM primary, obtaining AKM primaries in said representation provides a faster way to write down the null combinations than to diagonalize norm matrices.

Finally we must point out that there exist operators that are not AKM descendants and can never take part in an AKM primary combination. We already encountered such an example, namely the stress tensor: since it is not of Sugawara type it cannot be an AKM descendant, and the requirement that the AKM currents are Virasoro primaries implies that it also is not an AKM primary. Since the only other dimension two singlets are given by the quadratic Casimir operators, which have zero OPEs with the currents, one concludes that it is impossible to make an AKM primary combination involving the stress tensor. Another example of an operator which cannot be involved in any AKM primary combination is $(W \widetilde{W})\vert_{\mathrm{sing.}}$. We expect that the existence of this operator, as well as $(V V)\vert_{\mathrm{sing.}}$ is precisely the reason why the $T_4$ chiral algebra does not require (Virasoro primary) singlet generators of dimension three and four to close. Although these operators are not Virasoro primaries on their own, they take part in Virasoro primary combinations, of dimensions three and four respectively, which are not AKM primaries. Note that by being neither AKM primaries nor descendants, their contribution to the partition function is necessarily encrypted in the structure constant. As explained in section \ref{section_index}, their contribution is indeed captured by the $\mathrm{P.E.}\left[\frac{q^3+q^4}{1-q}\right]$ factor in the $T_4$ structure constant (see \eqref{structrureconst}). 

Looking at these operators it is natural to ask if the stress tensor and the Virasoro primary singlet operators obtained from  $(W \widetilde{W})\vert_{\mathrm{sing.}}$ and $(V V)\vert_{\mathrm{sing.}}$ form a closed subalgebra. If such an algebra closes, it must correspond to the $\mathcal{W}_4$ algebra, which is the unique (up to the choice of central charge) closed algebra with such a set of generators \cite{Kausch:1990bn,Blumenhagen:1990jv}. In principle this could be checked using our explicit construction; however, it is computationally challenging and we have not pursued it. More generally, one could wonder whether the set of operators $\mathcal{O}_i$ in Conjecture \ref{Tnconjecture} could form a closed subalgebra, which then should be a $\mathcal{W}_n=\mathcal{W}(2,3,\ldots,n)$ algebra with central charge $c_{2d}=-2n^3 + 3n^2 + n - 2$. In particular, one  should also be able to test this statement for $\chi(T_3)$ using the explicit construction of \cite{Beem:2014rza}, in which case one would obtain the  $\mathcal{W}_3$ algebra of \cite{Zamolodchikov:1985wn}.\\

In section \ref{section_index} we argued that the structure constant factor of \eqref{negativemodes} would add negative modes of the current algebra Casimir operators. Now we can give an explicit example: the operator $(J^1)^{b_1}_{a_1}W_{b_1 a_2a_3}$, which precisely at the critical level becomes an AKM primary, and thus is not included in the critical module of $W_{a_1 a_2 a_3}$. Taking the OPE of $S^1$ with $W_{a_1 a_2 a_3}$ we find
\begin{align}
S^1(z) W_{a_1 a_2 a_3}(0) &\sim \frac{15}{4} \frac{ W_{a_1 a_2 a_3}}{z^2} + 2 \frac{(J_1)_{a_1}^{b_1} W_{b_1 a_2 a_3}}{z}\,,  \nn \\
\Longleftrightarrow \left[ (S^1)_m, (W_{a_1 a_2 a_3})_n \right] &= \frac{15(m+1)}{4} (W_{a_1 a_2 a_3})_{m+n}  + 2 \left( (J_1)_{a_1}^{b_1} W_{b_1 a_2 a_3}\right)_{m+n}\,,
\end{align}
where $(\mathcal{O})_n$ denote the modes of operator $\mathcal{O}$, which in the case of $\left( (J_1)_{a_1}^{b_1} W_{b_1 a_2 a_3}\right)_{m+n}$ correspond to the modes of the normal-ordered product. Acting with the $(S_1)_{-1}$ mode of $S_1$ on the AKM primary yields
\be
(S^1)_{-1}|W_{a_1 a_2 a_3}\rangle   = (S^1)_{-1}(W_{a_1 a_2 a_3})_{-\frac{3}{2}}|0\rangle = 2 \left( (J_1)_{a_1}^{b_1} W_{b_1 a_2 a_3}\right)_{-\frac{5}{2}} |0\rangle\,,
\label{JWCas}
\ee
which exactly adds  $(J^1)^{b_1}_{a_1}W_{b_1 a_2a_3}$.\footnote{Recalling that the first null relation in \eqref{JW^(k) nulls}, sets equal $(J^1)^{b_1}_{a_1}W_{b_1 a_2a_3} = (J^2)^{b_2}_{a_2}W_{a_1 b_2a_3} = (J^3)^{b_3}_{a_3}W_{a_1 a_2b_3},$ this term and $L_{-1} |W_{a_1 a_2 a_3}\rangle$ (which produces $\partial W_{a_1 a_2 a_3}$) account for all the powers of $q$ in the structure constants, since for the fundamental representation only $2q$ survives in the plethystic exponential after combining \eqref{negativemodes} and \eqref{nullsfornegativemodes}. It can be shown that the OPEs of higher dimensional Casimirs with $W_{a_1 a_2 a_3}$ do not produce anything new.}

When analyzing the superconformal index we also argued that threefold AKM primaries in the sum over AKM modules, that do not correspond to generators $W^{(k)}$ must be obtained by normal-ordered products of generators of Higgs branch chiral ring origin. We can now give explicit examples. Let us start by considering representation $(\mathbf {15},\mathbf {15},\mathbf {15})$, for which the corresponding primary must have dimension three. As described in the previous section we can always write down a composite operator with the right quantum numbers, in this case it is $W \widetilde{W}\vert_{(\mathbf {15},\mathbf {15},\mathbf {15})}$. Even though this operator is not a threefold AKM primary, the following combination is:
\be 
W \widetilde{W}\vert_{(\mathbf {15},\mathbf {15},\mathbf {15})} + \frac{1}{64} (J^1)(J^2)(J^3)\,,
\label{151515AKM}
\ee
and it is precisely this combination that is accounted for by the $\mathfrak{R}=\mathbf{15}$ term in \eqref{partitionfunction}. Other examples at dimension three correspond to $(\mathbf {10},\mathbf {10},\mathbf {10})$ (and its conjugate), in which case the threefold AKM primary is simply $W W\vert_{(\mathbf {10},\mathbf {10},\mathbf {10})}$ (and $\widetilde{W} \widetilde{W} \vert_{(\overline{\mathbf{10}},\overline{\mathbf{10}},\overline{\mathbf{10}})}$).

\section*{Acknowledgments}
We are grateful to Leonardo Rastelli for suggesting the project and for guidance.
We would like to thank Chris Beem, Leonardo Rastelli and Balt van Rees for insightful discussions, and for comments on the manuscript. The work of ML is supported in part by FCT - Portugal through grant SFRH/BD/70614/2010. ML and WP are partially supported by NSF Grant PHY-1316617.

\appendix 

\section{Affine critical characters and the Schur index}
\label{App:character}

We show how to re-write the superconformal index \cite{Kinney:2005ej} in the so-called Schur limit \cite{Gadde:2011ik,Gadde:2011uv} in terms of characters of affine Kac-Moody modules at the critical level. The superconformal index of class $\mathcal{S}$ theories was computed in \cite{Gadde:2009kb,Gadde:2011ik,Gadde:2011uv,Gaiotto:2012xa}, and the characters of affine Kac-Moody algebras at the critical level in \cite{2007arXiv0706.1817A}. Here we just collect the final expressions and refer the readers to the original work for details.

Our conventions for affine Lie algebras follow those of \cite{DiFrancesco:1997nk}, and here we simply review some notation needed to write the characters. We denote the affine Lie algebra obtained by adding an imaginary root $\delta$ to a finite Lie algebra $\mathfrak{g}$ (of rank $r$) by $\hat{\mathfrak{g}}$.
The Cartan subalgebra of $\hat{\mathfrak{g}}$ ($\mathfrak{g}$) is denoted by $\hat{\mathfrak{h}}$ ($\mathfrak{h}$), and the positive roots of $\hat{\mathfrak{g}}$ ($\mathfrak{g}$) by $\hat{\Delta}_+$ ($\Delta_+$). 
We also denote the real positive roots of the affine Lie algebra, that is positive roots not of the form $n \delta$, by $\hat{\Delta}_+^{\mathrm{re}}$. 
The character of a critical irreducible highest weight representation $\mathfrak{R}_\lambda$ with highest weight $\hat{\lambda}$, whose restriction to the finite Lie algebra $\lambda$  is by definition an integral dominant weight is given in \cite{2007arXiv0706.1817A}. It reads \footnote{Here we used that for a critical highest weight $\hat{\lambda} + \hat{\rho} = \lambda + \rho$, and normalized the character to match the standard conventions for a partition function.}
\be 
\mathrm{ch}_{\mathfrak{R}_\lambda} = \frac{\sum_{w \in W} \epsilon(w) e^{ w \left( \lambda + \rho\right) - \rho } }{ \prod_{\alpha \in \Delta_+} \left( 1- q^{ \langle \lambda + \rho, \alpha^\vee \rangle }\right) \prod_{\hat{\alpha} \in \hat{\Delta}_+^{\mathrm{re}}} \left( 1- e^{-\hat{\alpha}}\right)}\,,
\label{critical_char}
\ee
where $W$ is the Weyl group of $\mathfrak{g}$, $\epsilon(w)$ is the signature of $w$, $q = e^{- \delta}$, $\rho$ denotes the Weyl vector, $\langle \cdot , \cdot \rangle$ denotes the Killing inner product and $\alpha^\vee$ is the coroot associated to $\alpha.$\\

\noindent
The Schur limit of the superconformal index of a $T_n$ theory is given by \cite{Gadde:2011ik,Gadde:2011uv}
\be 
\mathcal{I}_{T_n}(q; \mathbf x_i) =\sum_{\mathfrak{R}_\lambda} \frac{\prod_{i=1}^{3} \mathcal{K}_{\Lambda}(q;\mathbf{x}_i) \ \chi_{\mathfrak{R}_\lambda} (\mathbf{x}_i)}{\mathcal{K}_{\Lambda^t}(q)\  \text{dim}_q \mathfrak{R}_\lambda }\,,
\label{index}
\ee
with
\be 
\mathcal{K}_{\Lambda^t}(q) = \mathrm{P.E.}\left[\sum_{j=1}^{n-1}\frac{q^{d_j}}{1-q}\right]\,, \qquad 
\mathcal{K}_{\Lambda}(q;\mathbf{x})  = \mathrm{P.E.}\left[\frac{q \, \chi_\mathrm{adj.}(\mathbf{x)}}{1-q}\right]\,.
\ee
Here $\mathbf{x}_i$ denotes flavor fugacities conjugate to the Cartan generators of the $\mathfrak{su}(n)_i$ flavor group associated with each of the three punctures, $\Lambda$ and $\Lambda^t$ are respectively the trivial and principal embeddings of $\mathfrak{su}(2) \hookrightarrow \mathfrak{su}(n),$ and $d_j$ are the degrees of invariants. Furthermore, $\text{dim}_q \mathfrak{R}_\lambda$ is the $q$-deformed dimension of the representation $\mathfrak R_\lambda,$ \ie{}, 
\begin{equation}
\text{dim}_q \mathfrak{R}_\lambda = \prod_{\alpha \in \Delta_+} \frac{\left[\langle \lambda+\rho , \alpha \rangle\right]_q}{\left[\langle \rho , \alpha \rangle\right]_q}\,, \qquad\quad \text{where}\qquad [x]_q = \frac{q^{-\frac{x}{2}}-q^{\frac{x}{2}}}{q^{-\frac{1}{2}}-q^{\frac{1}{2}}} \,.
\label{qdim}
\end{equation} \\

\noindent
As shown in \cite{Beem:2014kka}, if $\lambda=0$ is the highest weight of the vacuum module 
\be 
\mathrm{ch}_{\mathfrak{R}_{\lambda=0}} = \frac{\mathcal{K}_{\Lambda}(q;\mathbf{x}_i)}{\mathcal{K}_{\Lambda^t}(q)}\,.
\ee
The expectation is that the full index for $T_n$ can be re-written as a sum of characters of critical modules\footnote{Along similar lines, one can rewrite the Schur limit of the superconformal index of the $T_{SO(2n)}$ theory \cite{Mekareeya:2012tn,Lemos:2012ph} in terms of critical affine $\widehat{\mathfrak{so}(2n)}$ characters.}. Re-writing \eqref{critical_char} to make manifest the vacuum module we find
\begin{equation}
\mathrm{ch}_{\mathfrak{R}_\lambda} = \mathrm{ch}_{\mathfrak{R}_{\lambda=0}}
\prod_{\alpha \in \Delta_+} \left( \frac{  1- q^{ \langle \rho , \alpha^\vee \rangle } }{  1- q^{ \langle \lambda + \rho , \alpha^\vee \rangle }} \right)
\frac{\sum_{w \in W} \epsilon(w) e^{ w \left(\lambda + \rho\right) - \rho  } }{\sum_{w \in W} \epsilon(w) e^{ w \left( \rho\right) - \rho  }}\,,
\end{equation}
where we recognize the last term as the character of the representation with highest weight $\lambda$ of $\mathfrak{g}$,
\be
\chi_\lambda(\mathbf{x}) =  \frac{\sum_{w \in W} \epsilon(w) e^{ w \left( \lambda + \rho \right)  } }{\sum_{w \in W} \epsilon(w) e^{ w \left( \rho\right)  }}\,.
\ee
After factoring out a $q^{-\langle \lambda,\rho \rangle},$ the middle factor can be written in terms of the $q$-deformed dimension \eqref{qdim} of the same representation:
\be
 \prod_{\alpha \in \Delta_+} \left( \frac{  1- q^{ \langle \rho , \alpha^\vee \rangle } }{  1- q^{ \langle \lambda + \rho , \alpha^\vee \rangle }} \right)= 
 \prod_{\alpha \in \Delta_+} q^{-\langle \lambda , \alpha^\vee \rangle /2}  \prod_{\alpha \in \Delta_+}  \left( \frac{ q^{ -\langle \rho , \alpha^\vee \rangle/2 } - q^{ \langle \rho , \alpha^\vee \rangle /2} }{  q^{ -\langle \lambda + \rho , \alpha^\vee \rangle/2 } - q^{ \langle \lambda + \rho , \alpha^\vee \rangle/2 }}\right) = q^{-\langle \lambda, \rho \rangle} \frac{1}{\dim_q \mathfrak{R}_\lambda} \,,
\ee
where we used that $\alpha^\vee = \alpha$, for $\mathfrak{su}(n)$, to identify $\rho$ in the last step. In total we thus find
\begin{equation}\label{explicitcriticalcharacter}
\mathrm{ch}_{\mathfrak{R}_\lambda} = \frac{\mathrm{P.E.}\left[\frac{q \, \chi_\mathrm{adj.}(\mathbf{x)}}{1-q}\right] \chi_\lambda(\mathbf{x})  }{ q^{\langle \lambda, \rho \rangle} \mathrm{P.E.}\left[ \sum_{j=1}^{n-1}\frac{q^{d_j}}{1-q}\right] \dim_q \mathfrak{R}_\lambda}\,.
\end{equation}
Using this result in the expression for the superconformal index \eqref{index} we obtain \eqref{partitionfunction}. To obtain \eqref{structrureconst} we also note that the denominator of \eqref{explicitcriticalcharacter} can be rewritten as
\begin{align}
q^{\langle \lambda, \rho \rangle}\;  \mathrm{P.E.}\left[\sum_{j=1}^{n-1}\frac{q^{d_j}}{1-q}\right]\; \text{dim}_q \mathfrak{R}_\lambda
&= \mathrm{P.E.}\left[ \sum_{j=1}^{n-1} \frac{q^{d_j}}{1-q} + \sum_{\alpha\in \Delta_+} q^{\langle \rho, \alpha \rangle} - \sum_{\alpha\in \Delta_+} q^{\langle \lambda + \rho, \alpha \rangle} \right] \notag\\
&= \mathrm{P.E.}\left[ \sum_{j=1}^{n-1} \frac{q^{d_j}}{1-q} + \sum_{j=1}^{n-1} \left( n-j \right) q^j - \sum_{j=2}^{n}\sum_{1\leq i < j} q^{\ell_i - \ell_j + j-i} \right]\,.
\end{align}

\section{The OPEs}
\label{App:OPE}
In this appendix we give all the OPEs between the generators of the $T_4$ chiral algebra. Here all OPE coefficients (including the central charges) are already set to the values required by the Jacobi-identities, as described in Section \ref{section_T4}. Since all generators are both Virasoro and AKM primaries, with the exception of the stress tensor which is neither and the AKM currents which are not AKM primaries, all singular OPEs involving the affine currents and the stress tensor are completely fixed by flavor symmetries and Virasoro symmetry, up to the flavor central charges $(k_{2d})_{i=1,2,3} = -4$ and the Virasoro central charge $c_{2d}= -78$ appearing in the most singular term in their respective self-OPEs. Different AKM currents are taken to have zero singular OPE.  As discussed in Section~\ref{section_T4} we consistently treat the three flavor symmetries on equal footing, in particular we require $k_{2d}\equiv(k_{2d})_{1}=(k_{2d})_{2}=(k_{2d})_{3}$ . We recall that also the precise values of $c_{2d}$ and $k_{2d}$ central charges are a result of imposing the Jacobi-identities.

The singular OPEs of the $W,\widetilde W$ generators among themselves were found to be
\begin{align*}
W_{a_1a_2a_3}(z)\  W_{b_1b_2b_3}(0)\  \sim\  & \frac{1}{2\,z} V_{[a_1b_1][a_2b_2][a_3b_3]}\,,\\
\widetilde W^{a_1a_2a_3}(z)\  \widetilde W^{b_1b_2b_3}(0) \ \sim \  & \frac{1}{2\,z} \frac{1}{8} \epsilon^{a_1b_1c_1d_1}\epsilon^{a_2b_2c_2d_2}\epsilon^{a_3b_3c_3d_3}V_{[c_1d_1][c_2d_2][c_3d_3]}\,,\\
W_{a_1a_2a_3}(z)\  \widetilde W^{b_1b_2b_3}(0) \ \sim \  & \frac{1}{z^3} \delta^{b_1}_{a_1}\delta^{b_2}_{a_2}\delta^{b_3}_{a_3} - \frac{1}{4\,z^2} \left( \delta^{b_1}_{a_1}\delta^{b_2}_{a_2} (J^3)^{b_3}_{a_3} +\text{perms.}\right)\\
&-\frac{1}{4\,z} \left( \delta^{b_1}_{a_1}\delta^{b_2}_{a_2} \partial(J^3)^{b_3}_{a_3} +\text{perms.}\right) +\frac{1}{16\,z} \left( \delta^{b_1}_{a_1}(J^2)^{b_2}_{a_2} (J^3)^{b_3}_{a_3} +\text{perms.}\right)\\
&+\frac{1}{z}\delta^{b_1}_{a_1}\delta^{b_2}_{a_2}\delta^{b_3}_{a_3}\left(-\frac{1}{16} T -\frac{1}{96} \left((J^1)^{\alpha_1}_{\beta_1}(J^1)^{\beta_1}_{\alpha_1}+\text{2 more}\right)\right)\\
&+\frac{ 1 }{16\,z}\left(\delta^{b_1}_{a_1}\delta^{b_2}_{a_2} (J^3)^{\alpha_3}_{a_3}(J^3)^{b_3}_{\alpha_3}+\text{perms.}\right)\,,
\end{align*}
where we have fixed the normalization of $W$ and $\widetilde W$ to convenient values. In all these OPEs ``$+$2 more'' means we must add the same term for the remaining two currents, and ``$+$perms.'' that all independent permutations of the previous term must be added. We also found the OPEs between the $W,\widetilde W$ and $V$ generators to be
\begin{align*}
W_{a_1a_2a_3}(z)\ V_{[b_1c_1][b_2c_2][b_3c_3]}(0) \ \sim \ & \frac{1}{8} \epsilon_{a_1b_1c_1d_1}\epsilon_{a_2b_2c_2d_2}\epsilon_{a_3b_3c_3d_3} \left(-\frac{3}{z^2}\widetilde W^{d_1d_2d_3} - \frac{1}{z} \partial\widetilde W^{d_1d_2d_3} \right)\\
& -\frac{1}{8}\epsilon_{a_1b_1c_1d_1}\epsilon_{a_2b_2c_2d_2}\epsilon_{a_3b_3c_3d_3}\frac{1}{3\,z}\left( (J^1)^{d_1}_{\alpha_1}\widetilde W^{\alpha_1d_2d_3} + \text{perms.}\right)  \\
&-\frac{1}{8\,z}\left( \epsilon_{\alpha_1b_1c_1d_1}\epsilon_{a_2b_2c_2d_2}\epsilon_{a_3b_3c_3d_3}(J^1)^{d_1}_{a_1}\widetilde W^{\alpha_1d_2d_3} + \text{perms.}\right)\,, \\
\widetilde W^{a_1a_2a_3}(z)\ V_{[b_1c_1][b_2c_2][b_3c_3]}(0) \ \sim \ &  \delta^{a_1}_{[b_1}\delta^{a_2}_{[b_2}\delta^{a_3}_{[b_3} \left(\frac{3}{z^2}W_{c_1]c_2]c_3]} +  \frac{1}{z} \partial W_{c_1]c_2]c_3]}\right)\\
& +\frac{1}{3\,z} \delta^{a_1}_{[b_1}\delta^{a_2}_{[b_2}\delta^{a_3}_{[b_3} \left( (J^1)_{c_1]}^{\alpha_1} W_{\alpha_1c_2]c_3]} + \text{perms.}\right)  \\
&+\frac{1}{z}\left(\delta^{a_2}_{[b_2}\delta^{a_3}_{[b_3}(J^1)^{a_1}_{[b_1} W_{c_1]c_2]c_3]} + \text{perms.}\right)\,.
\end{align*}
Finally, the singular $VV$ OPE reads
\begingroup
\allowdisplaybreaks[1]
\begin{align*}
&V_{[a_1b_1][a_2b_2][a_3b_3]}(z) V^{[c_1d_1][c_2d_2][c_3d_3]}(0) \\*
&\sim \ \delta^{[c_1}_{a_1}\delta^{d_1]}_{b_1} \delta^{[c_2}_{a_2}\delta^{d_2]}_{b_2}\delta^{[c_3}_{a_3}\delta^{d_3]}_{b_3}\left( \frac{6}{z^4}-\frac{1}{2\,z^2}T - \frac{1}{4\,z}\partial T -   \frac{1}{24\,z^2}\left((J^1)^{\alpha_1}_{\beta_1}(J^1)^{\beta_1}_{\alpha_1}+\text{2 more}\right) \right. \\*
&\phantom{\sim \ \delta^{[c_1}_{a_1}\delta^{d_1]}_{b_1} \delta^{[c_2}_{a_2}\delta^{d_2]}_{b_2}\delta^{[c_3}_{a_3}\delta^{d_3]}_{b_3}\quad} \left.
- \frac{19}{480\,z}\partial\left((J^1)^{\alpha_1}_{\beta_1}(J^1)^{\beta_1}_{\alpha_1}+\text{2 more}\right) - \frac{37}{20\,z} W_{\alpha_1\beta_1\gamma_1}\widetilde W^{\alpha_1\beta_1\gamma_1}  \right. \\*
&\phantom{\sim \ \delta^{[c_1}_{a_1}\delta^{d_1]}_{b_1} \delta^{[c_2}_{a_2}\delta^{d_2]}_{b_2}\delta^{[c_3}_{a_3}\delta^{d_3]}_{b_3}\quad} \left.
+  \frac{1}{40\,z} \left((J^1)_{\alpha_1}^{\beta_1}(J^1)_{\beta_1}^{\gamma_1}(J^1)^{\alpha_1}_{\gamma_1}+\text{2 more}\right) \right)\\
&\phantom{\sim \ }+\delta^{[c_2}_{a_2}\delta^{d_2]}_{b_2}\delta^{[c_3}_{a_3}\delta^{d_3]}_{b_3}\left(-\frac{3}{16\,z}\delta_{[b_1}^{[d_1}(J^1)_{a_1]}^{|\gamma_1} (J^1)^{\beta_1|}_{\gamma_1}(J^1)^{c_1]}_{\beta_1} + \frac{33}{80\,z} (J^1)^{\beta_1}_{[a_1}(J^1)^{[d_1}_{|\beta_1|}(J^1)_{b_1]}^{c_1]} - \frac{1}{4\,z} (J^1)^{[d_1}_{[a_1}(J^1)_{b_1]}^{c_1]} \right.\\*
&\phantom{\sim \ \delta^{[c_2}_{a_2}\delta^{d_2]}_{b_2}\delta^{[c_3}_{a_3}\delta^{d_3]}_{b_3}\quad}\left.-\frac{43}{80\,z^2} \partial\left((J^1)^{[d_1}_{[a_1}(J^1)_{b_1]}^{c_1]}\right) -\frac{3}{2\,z^3} (J^1)_{[a_1}^{[c_1}\delta_{b_1]}^{d_1]} + \frac{1}{40\,z^3} T(J^1)_{[a_1}^{[c_1}\delta_{b_1]}^{d_1]} \right.\\*
&\phantom{\sim \ \delta^{[c_2}_{a_2}\delta^{d_2]}_{b_2}\delta^{[c_3}_{a_3}\delta^{d_3]}_{b_3}\quad}\left.- \frac{11}{120\,z} \left( (J^1)^{\alpha_1}_{\beta_1}(J^1)^{\beta_1}_{\alpha_1}+\text{2 more} \right)(J^1)_{[a_1}^{[c_1}\delta_{b_1]}^{d_1]} - \frac{5}{4\,z^2} \partial(J^1)_{[a_1}^{[c_1}\delta_{b_1]}^{d_1]}   \right.\\*
&\phantom{\sim \ \delta^{[c_2}_{a_2}\delta^{d_2]}_{b_2}\delta^{[c_3}_{a_3}\delta^{d_3]}_{b_3}\quad}\left.+ \frac{1}{4\,z^2} (J^1)^{\alpha_1}_{[a_1}(J^1)_{|\alpha_1|}^{[c_1}\delta^{d_1]}_{b_1]}+ \frac{17}{40\,z} \partial(J^1)^{\alpha_1}_{[a_1}(J^1)_{|\alpha_1|}^{[c_1}\delta^{d_1]}_{b_1]}+ \frac{7}{40\,z} (J^1)^{\alpha_1}_{[a_1}\partial(J^1)_{|\alpha_1|}^{[c_1}\delta^{d_1]}_{b_1]}\right.\\*
&\phantom{\sim \ \delta^{[c_2}_{a_2}\delta^{d_2]}_{b_2}\delta^{[c_3}_{a_3}\delta^{d_3]}_{b_3}\quad}\left.-\frac{23}{80\,z} \partial^2(J^1)_{[a_1}^{[c_1}\delta_{b_1]}^{d_1]}\right)\\*
&\qquad\qquad + \text{permutations}[1,2,3]\\
&\phantom{\sim \ }+\delta^{[c_3}_{a_3}\delta^{d_3]}_{b_3}\left(\frac{1}{4	\,z} \delta_{[b_2}^{[d_2}(J^2)_{a_2]}^{c_2]}(J^1)_{[a_1}^{[d_1}(J^1)_{b_1]}^{c_1]}  + \frac{3}{4\,z^2} \delta_{[b_2}^{[d_2}(J^2)_{a_2]}^{c_2]}(J^1)_{[a_1}^{[c_1}\delta_{b_1]}^{d_1]}+ 
\right. \\*
&\phantom{\sim \ +\delta^{[c_3}_{a_3}\delta^{d_3]}_{b_3}\quad}\left.+\frac{13}{4\,z} \delta_{[b_2}^{[d_2}(J^2)_{a_2]}^{c_2]}\partial(J^1)_{[a_1}^{[c_1}\delta_{b_1]}^{d_1]} -\frac{3}{4\,z} \delta_{[b_2}^{[d_2}(J^2)_{a_2]}^{c_2]}(J^1)_{[a_1}^{\alpha_1}(J^1)_{|\alpha_1|}^{c_1]}\delta_{b_1]}^{d_1}  \right)\\*
&\qquad\qquad + \text{permutations}[1,2,3]\\
&\phantom{\sim \ } + \frac{19}{5\,z}\left(\delta^{[c_2}_{a_2}\delta^{d_2]}_{b_2}\delta^{[c_3}_{a_3}\delta^{d_3]}_{b_3}\delta_{[b_1}^{[d_1}W_{a_1]\beta_2\gamma_3}\widetilde W^{c_1]\beta_2\gamma_3} + \delta^{[c_3}_{a_3}\delta^{d_3]}_{b_3}\delta^{[c_1}_{a_1}\delta^{d_1]}_{b_1}\delta_{[b_2}^{[d_2}W_{\beta_1a_2]\gamma_3}\widetilde W^{\beta_1c_2]\gamma_3} \right. \\*
&\phantom{\sim \  + \frac{19}{5\,z}\quad} \left. +\delta^{[c_1}_{a_1}\delta^{d_1]}_{b_1}\delta^{[c_2}_{a_2}\delta^{d_2]}_{b_2}\delta_{[b_3}^{[d_3}W_{\beta_1\gamma_2a_3]}\widetilde W^{\beta_1\gamma_2c_3]} \right)\\
&\phantom{\sim \ } - \frac{4}{z}\left( \delta^{[c_3}_{a_3}\delta^{d_3]}_{b_3}\delta_{[b_1}^{[d_1} \delta_{[b_2}^{[d_2} W_{a_1]a_2]\gamma_3}\widetilde W^{c_1]c_2]\gamma_3} + \delta^{[c_1}_{a_1}\delta^{d_1]}_{b_1}\delta_{[b_2}^{[d_2} \delta_{[b_3}^{[d_3} W_{\gamma_1a_2]a_3]}\widetilde W^{\gamma_1c_2]c_3]} \right. \\*
&\phantom{\sim \  - \frac{4}{z}\quad} \left.+ \delta^{[c_2}_{a_2}\delta^{d_2]}_{b_2}\delta_{[b_3}^{[d_3} \delta_{[b_1}^{[d_1} W_{a_1]\gamma_2a_3]}\widetilde W^{c_1]\gamma_2c_3]}   \right)\\
&\phantom{\sim \ } - \frac{1}{z} \delta_{[b_1}^{[d_1} \delta_{[b_2}^{[d_2} \delta_{[b_3}^{[d_3} (J^1)_{a_1]}^{c_1]}(J^2)_{a_2]}^{c_2]}(J^3)_{a_3]}^{c_3]}  - \frac{16}{z} \delta_{[b_1}^{[d_1} \delta_{[b_2}^{[d_2} \delta_{[b_3}^{[d_3} W_{a_1]a_2]a_3]}\widetilde W^{c_1]c_2]c_3]}\,,
\end{align*}
\endgroup
where the norm of $V$ was also fixed, and for convenience we defined $V^{[c_1d_1][c_2d_2][c_3d_3]}$ through $V_{[a_1 b_1] [a_2 b_2] [a_3 b_3]}= \frac{1}{8}\epsilon_{a_1 b_1 c_1 d_1}\epsilon_{a_2 b_2 c_2 d_2}\epsilon_{a_3 b_3 c_3 d_3} V^{[c_1d_1][c_2d_2][c_3d_3]}$. Here ``$\text{permutations}[1,2,3]$'' means we must repeat the previous term with all possible permutations of the flavor groups indices.

{
\bibliographystyle{utphys}
\bibliography{chiralTN}
}

\end{document}